\documentclass{article}
\usepackage{spconf,amsmath,graphicx}
\usepackage{amssymb,bm,booktabs,tabularx,url}
\usepackage[colorlinks]{hyperref}
\newcolumntype{C}{>{\centering\arraybackslash}X}
\newcolumntype{L}{>{\raggedright\arraybackslash}X}
\newcolumntype{R}{>{\raggedleft\arraybackslash}X}
\makeatletter
\newcommand\footnoteref[1]{\protected@xdef\@thefnmark{\ref{#1}}\@footnotemark}
\makeatother

\title{MASKCYCLEGAN-VC:
  \\LEARNING NON-PARALLEL VOICE CONVERSION WITH FILLING IN FRAMES}
\name{Takuhiro Kaneko, Hirokazu Kameoka, Kou Tanaka, Nobukatsu Hojo}
\address{NTT Communication Science Laboratories, NTT Corporation, Japan}

\begin{document}
\maketitle
\begin{abstract}
  Non-parallel voice conversion (VC) is a technique for training voice converters without a parallel corpus. Cycle-consistent adversarial network-based VCs (CycleGAN-VC and CycleGAN-VC2) are widely accepted as benchmark methods. However, owing to their insufficient ability to grasp time-frequency structures, their application is limited to mel-cepstrum conversion and not mel-spectrogram conversion despite recent advances in mel-spectrogram vocoders. To overcome this, CycleGAN-VC3, an improved variant of CycleGAN-VC2 that incorporates an additional module called time-frequency adaptive normalization (TFAN), has been proposed. However, an increase in the number of learned parameters is imposed. As an alternative, we propose MaskCycleGAN-VC, which is another extension of CycleGAN-VC2 and is trained using a novel auxiliary task called filling in frames (FIF). With FIF, we apply a temporal mask to the input mel-spectrogram and encourage the converter to fill in missing frames based on surrounding frames. This task allows the converter to learn time-frequency structures in a self-supervised manner and eliminates the need for an additional module such as TFAN. A subjective evaluation of the naturalness and speaker similarity showed that MaskCycleGAN-VC outperformed both CycleGAN-VC2 and CycleGAN-VC3 with a model size similar to that of CycleGAN-VC2.\footnote{\label{foot:samples}Audio samples are available at \url{http://www.kecl.ntt.co.jp/people/kaneko.takuhiro/projects/maskcyclegan-vc/index.html}.}
\end{abstract}


\section{Introduction}
\label{sec:introduction}

Voice conversion (VC) is a technique for translating one voice into another without changing the linguistic content, and has been extensively studied owing to its various applications, including speaking assistance~\cite{AKainSC2007,KNakamuraSC2012}, speech enhancement~\cite{ZInanogluSC2009,TTodaTASLP2012}, and accent conversion~\cite{DFelpsSC2009,TKanekoIS2017b}.
Machine-learning-based approaches have been widely used, ranging from statistical modeling (e.g., Gaussian mixture models~\cite{YStylianouTASP1998,TTodaTASLP2007}) to neural networks (NNs) (e.g., feedforward NNs~\cite{SDesaiTASLP2010}, recurrent NNs~\cite{LSunICASSP2015}, convolutional NNs~\cite{TKanekoIS2017b}, and attention networks~\cite{JXZhangTASLP2019,KTanakaICASSP2019,HKameokaTASLP2020}).

Many VC methods (including those above) are categorized into parallel VC approaches and train a converter between the source and target speakers using parallel utterances.
Parallel VC has the advantage that it can train a converter in a supervised manner; however, it requires a parallel corpus, which is not always easy to collect.

As an alternative, non-parallel VC, a technique for training a converter without a parallel corpus, has attracted attention, and many such methods have thus been proposed.
Among them, a promising approach is to utilize linguistic information to compensate for the missing parallel supervision~\cite{LSunICME2016,FLXieIS2016,YSaitoICASSP2018,JZhangTASLP2020}; however, extra data or pretrained models are needed to derive such linguistic information.

To remove such a requirement and solve non-parallel VC without any additional data or pretrained models, deep generative models, such as generative adversarial networks (GANs)~\cite{IGoodfellowNIPS2014} and variational autoencoders (VAEs)~\cite{DKingmaICLR2014}, have been introduced~\cite{CHsuAPSIPA2016,CHsuIS2017,TKanekoArXiv2017,HKameokaTASLP2019,PTobingIS2019}.
Among them, the family of CycleGAN-VCs (CycleGAN-VC~\cite{TKanekoArXiv2017,TKanekoEUSIPCO2018}, CycleGAN-VC2~\cite{TKanekoICASSP2019}, and StarGAN-VCs~\cite{HKameokaSLT2018,TKanekoIS2019,HKameokaTASLP2020b}) are significant achievements and have been widely accepted as benchmark approaches (e.g., \cite{JZhangTASLP2020,SLeeICASSP2020,KQianICML2019}).
However, owing to their insufficient capacity to capture the time-frequency structure (e.g., the harmonic structure is compromised, as shown in Figure 1 in~\cite{TKanekoIS2020}), their application is limited to mel-cepstrum conversion and not mel-spectrogram conversion despite recent advances in mel-spectrogram vocoders~\cite{JShenICASSP2018,RPrengerICASSP2019,KKumarNeurIPS2019,RYamamotoICASSP2020,NChenArXiv2020}.

To overcome this, CycleGAN-VC3~\cite{TKanekoIS2020}, an improved variant of CycleGAN-VC2, was recently proposed, and addresses the problem by incorporating an additional module called time-frequency adaptive normalization (TFAN).
Although the performance is superior, an increase in the number of converter parameters is necessary (from 16M to 27M).

As an alternative, we propose \textit{MaskCycleGAN-VC}, which is another extension of CycleGAN-VC2 and is trained using a novel auxiliary task called \textit{filling in frames (FIF)}.
With FIF, we apply a temporal mask to the input mel-spectrogram and encourage the converter to fill in the missing frames based on the surrounding frames.
FIF is inspired by the success of complementation-based self-supervised learning in other fields, e.g., image inpainting in computer vision~\cite{DPathakCVPR2016} and text infilling in natural language processing~\cite{WFedusICLR2018,JDevlinNAACL2019}.
Similarly, FIF allows the converter to learn the time-frequency feature structure in a self-supervised manner through a complementation process.
This strong property eliminates the need for an additional module such as TFAN, and makes CycleGAN-VC2 applicable to mel-spectrogram conversion with negligibly small network modifications.

We investigated the effectiveness of MaskCycleGAN-VC on the Spoke (i.e., non-parallel VC) task of the Voice Conversion Challenge 2018 (VCC 2018)~\cite{VCC2018}.
A subjective evaluation of the naturalness and speaker similarity showed that MaskCycleGAN-VC outperformed both CycleGAN-VC2 and CycleGAN-VC3 while keeping the model size similar to that of CycleGAN-VC2.

The rest of this paper is organized as follows.
In Section~\ref{sec:cyclegan-vc2}, we review CycleGAN-VC2, which is the baseline of our model.
We then introduce the proposed MaskCycleGAN-VC in Section~\ref{sec:maskcyclegan-vc}.
In Section~\ref{sec:experiments}, we describe the experimental results.
Finally, we provide some concluding remarks and areas of future study in Section~\ref{sec:conclusions}.

\section{Conventional CycleGAN-VC2}
\label{sec:cyclegan-vc2}

The purpose of CycleGAN-VC2 is to train a converter $G_{X \rightarrow Y}$ that translates source acoustic features $\bm{x} \in X$ into target acoustic features $\bm{y} \in Y$ without parallel supervision.
Following CycleGAN~\cite{JYZhuICCV2017,ZYiICCV2017,TKimICML2017}, which was proposed for unpaired image-to-image translation, CycleGAN-VC2 solves this problem using an \textit{adversarial loss}~\cite{IGoodfellowNIPS2014}, \textit{cycle-consistency loss}~\cite{TZhouCVPR2016}, and \textit{identity-mapping loss}~\cite{YTaigmanICLR2017}.
In addition, CycleGAN-VC2 uses a \textit{second adversarial loss}~\cite{TKanekoICASSP2019} to improve the quality of the cyclically reconstructed features.

\smallskip\noindent\textbf{Adversarial loss.}
An adversarial loss $\mathcal{L}_{adv}^{X \rightarrow Y}$ is used to make the converted feature $G_{X \rightarrow Y}(\bm{x})$ appear to be the target:
\begin{flalign}
  \label{eqn:adv}
  \mathcal{L}_{adv}^{X \rightarrow Y} & = \mathbb{E}_{\bm{y} \sim P_Y} [\log D_Y(\bm{y})] \nonumber \\
  & + \mathbb{E}_{\bm{x} \sim P_X} [\log(1 - D_Y(G_{X \rightarrow Y}(\bm{x})))],
\end{flalign}
where the discriminator $D_Y$ distinguishes a real $\bm{y}$ from the generated $G_{X \rightarrow Y}(\bm{x})$ by maximizing this loss, whereas $G_{X \rightarrow Y}$ generates $G_{X \rightarrow Y}(\bm{x})$, which can deceive $D_Y$ by minimizing this loss.
Similarly, the inverse converter $G_{Y \rightarrow X}$ is trained with the discriminator $D_X$ using $\mathcal{L}_{adv}^{Y \rightarrow X}$.

\smallskip\noindent\textbf{Cycle-consistency loss.}
A cycle-consistency loss $\mathcal{L}_{cyc}^{X \rightarrow Y \rightarrow X}$ is used to determine the pseudo pair within the cycle-consistency constraint without parallel supervision:
\begin{flalign}
  \label{eqn:cyc}
  \mathcal{L}_{cyc}^{X \rightarrow Y \rightarrow X} = \mathbb{E}_{\bm{x} \sim P_X} [ \| G_{Y \rightarrow X}(G_{X \rightarrow Y}(\bm{x})) - \bm{x} \|_1 ].
\end{flalign}
Similarly, $\mathcal{L}_{cyc}^{Y \rightarrow X \rightarrow Y}$ is used for the inverse-forward mapping (i.e., $G_{X \rightarrow Y}(G_{Y \rightarrow X}(\bm{y}))$).

\smallskip\noindent\textbf{Identity-mapping loss.}
An identity-mapping loss $\mathcal{L}_{id}^{X \rightarrow Y}$ is used to enhance the input preservation:
\begin{flalign}
  \label{eqn:id}
  \mathcal{L}_{id}^{X \rightarrow Y} = \mathbb{E}_{\bm{y} \sim P_Y} [ \| G_{X \rightarrow Y}(\bm{y}) - \bm{y} \|_1 ].
\end{flalign}
Similarly, $\mathcal{L}_{id}^{Y \rightarrow X}$ is used for the inverse converter $G_{Y \rightarrow X}$.

\smallskip\noindent\textbf{Second adversarial loss.}
A second adversarial loss $\mathcal{L}_{adv2}^{X \rightarrow Y \rightarrow X}$ is used to mitigate the statistical averaging caused by L1 loss in Eq.~\ref{eqn:cyc}.
\begin{flalign}
  \label{eqn:adv2}
  &\mathcal{L}_{adv2}^{X \rightarrow Y \rightarrow X} = \mathbb{E}_{\bm{x} \sim P_X} [ \log D_{X}'(\bm{x}) ] \nonumber \\
  & \:\:\:\:\:\: + \mathbb{E}_{\bm{x} \sim P_X} [ \log (1 - D_{X}'(G_{Y \rightarrow X}(G_{X \rightarrow Y}(\bm{x})))) ],
\end{flalign}
where the discriminator $D_{X}'$ distinguishes a reconstructed $G_{Y \rightarrow X}(G_{X \rightarrow Y}(\bm{x}))$ from a real $\bm{x}$.
Similarly, $\mathcal{L}_{adv2}^{Y \rightarrow X \rightarrow Y}$ is used for the inverse-forward mapping with an additional discriminator $D_{Y}'$.

\smallskip\noindent\textbf{Full objective.}
A full objective $\mathcal{L}_{full}$ is written as follows:
\begin{flalign}
  & \mathcal{L}_{full} = \mathcal{L}_{adv}^{X \rightarrow Y} + \mathcal{L}_{adv}^{Y \rightarrow X} + \lambda_{cyc} (\mathcal{L}_{cyc}^{X \rightarrow Y \rightarrow X} + \mathcal{L}_{cyc}^{Y \rightarrow X \rightarrow Y}) \nonumber \\
  & \:\: + \lambda_{id} (\mathcal{L}_{id}^{X \rightarrow Y} + \mathcal{L}_{id}^{Y \rightarrow X}) + \mathcal{L}_{adv2}^{X \rightarrow Y \rightarrow X} + \mathcal{L}_{adv2}^{Y \rightarrow X \rightarrow Y},
\end{flalign}
where $\lambda_{cyc}$ and $\lambda_{id}$ are weighing parameters.
$G_{X \rightarrow Y}$ and $G_{Y \rightarrow X}$ are optimized by minimizing this loss, whereas $D_X$, $D_Y$, $D_{X}'$, and $D_{Y}'$ are optimized by maximizing this loss.

\begin{figure}[t]
  \centerline{\includegraphics[width=\columnwidth]{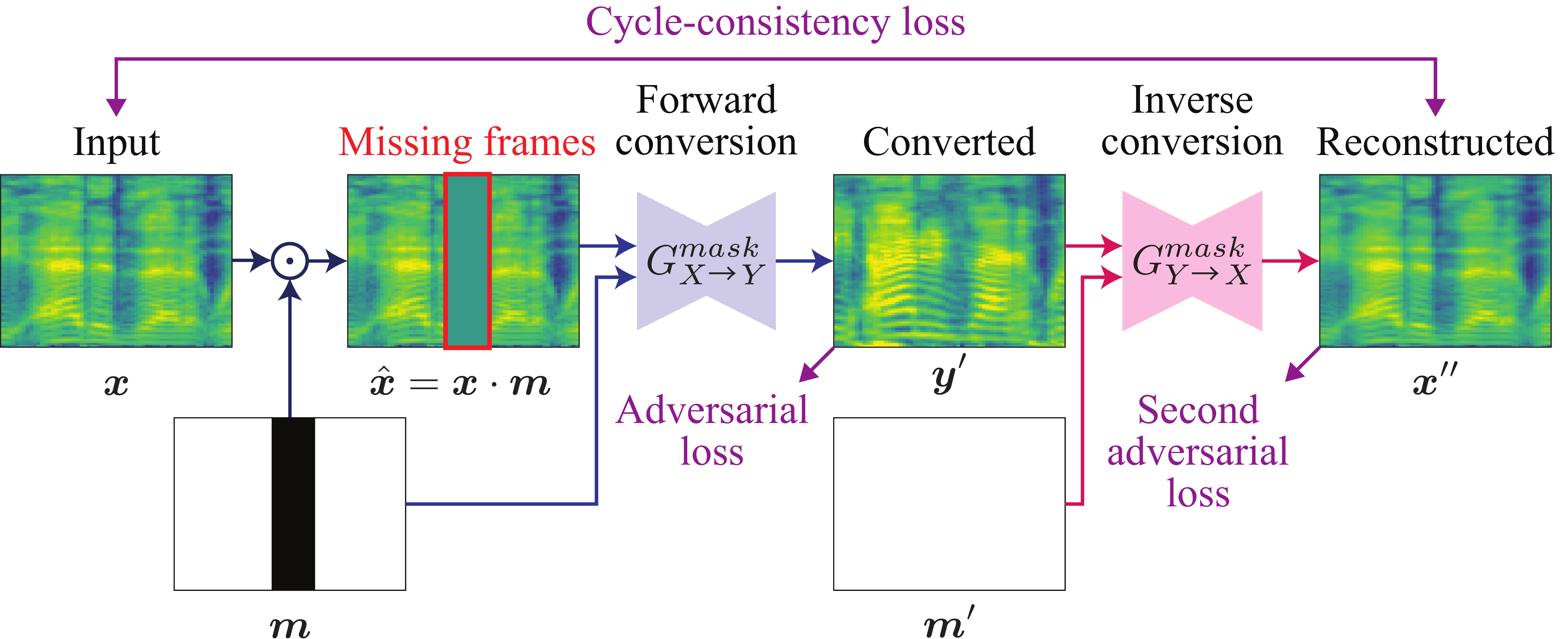}}
  \vspace{-3mm}
  \caption{Pipeline of FIF for the forward-inverse mapping.
    We encourage the converter to fill in the missing frames (surrounded by the red box) based on the surrounding frames through a cyclic conversion process.
    In practice, a similar procedure is used for the inverse-forward mapping.}
  \vspace{-3.5mm}
  \label{fig:pipeline}
\end{figure}

\section{MaskCycleGAN-VC}
\label{sec:maskcyclegan-vc}

\subsection{Training with Filling in Frames (FIF)}
\label{subsec:fif}

As shown in \cite{TKanekoIS2020}, CycleGAN-VC2, which was developed for mel-cepstrum conversion, does not have sufficient ability to capture the time-frequency structure in mel-spectrogram conversion; consequently, the harmonic structure is often compromised.
To alleviate this, we devised \textit{MaskCycleGAN-VC}, which is trained using the auxiliary \textit{FIF} task.
We present the overall pipeline of FIF in Fig.~\ref{fig:pipeline}.

Given the source mel-spectrogram $\bm{x}$, we first create a temporal mask $\bm{m} \in M$, which has the same size as $\bm{x}$, parts of which have a value of zero (denoted by the black region in Fig.~\ref{fig:pipeline}), and the remaining parts have a value of 1 (indicated by the white region in Fig.~\ref{fig:pipeline}).
A masked region (i.e., zero region) is randomly determined based on a predetermined rule (the effect of which is examined in Section~\ref{subsec:objective_evaluation}).

Subsequently, we apply the mask $\bm{m}$ to $\bm{x}$ as follows:
\begin{flalign}
  \label{eqn:mask}
  \hat{\bm{x}} = \bm{x} \cdot \bm{m},
\end{flalign}
where $\cdot$ represents an element-wise product.
By using this procedure, we artificially create missing frames, as shown in the region surrounded by the red box in Fig.~\ref{fig:pipeline}.

Next, the MaskCycleGAN-VC converter $G_{X \rightarrow Y}^{mask}$ synthesizes $\bm{y}'$ from $\hat{\bm{x}}$ and $\bm{m}$ as follows:
\begin{flalign}
  \label{eqn:mask_conversion}
  \bm{y}' = G_{X \rightarrow Y}^{mask}(\mathrm{concat} (\hat{\bm{x}}, \bm{m})),
\end{flalign}
where $\mathrm{concat}$ denotes the channel-wise concatenation.
By using $\bm{m}$ as the conditional information, $G_{X \rightarrow Y}^{mask}$ can fill in the frames while knowing which frames need to be filled in.

Similar to CycleGAN-VC2, we can ensure that $\bm{y}'$ is in the target $Y$ by using an adversarial loss (Eq.~\ref{eqn:adv}) but cannot compare $\bm{y}'$ with the ground truth directly owing to the lack of parallel supervision.
As an alternative, we aim to fill in the frames through a cyclic conversion process.
To do so, we reconstruct $\bm{x}''$ using the inverse converter $G_{Y \rightarrow X}^{mask}$:
\begin{flalign}
  \label{eqn:mask_reconstruction}
  \bm{x}'' = G_{Y \rightarrow X}^{mask}(\mathrm{concat}(\bm{y}', \bm{m}')),
\end{flalign}
where $\bm{m}'$ is represented using an all-ones matrix under the assumption that the missing frames have been filled in ahead of this process.
We then apply the cycle-consistency loss for the original and reconstructed mel-spectrograms:
\begin{flalign}
  \label{eqn:mask_cyc}
  \mathcal{L}_{mcyc}^{X \rightarrow Y \rightarrow X} = \mathbb{E}_{\bm{x} \sim P_X, \bm{m} \sim P_M} [ \| \bm{x}'' - \bm{x} \|_1 ],
\end{flalign}
where we simultaneously used a second adversarial loss (Eq.~\ref{eqn:adv2}) for $\bm{x}''$.

To optimize $\mathcal{L}_{mcyc}^{X \rightarrow Y \rightarrow X}$, $G_{X \rightarrow Y}^{mask}$ needs to derive information useful for filling in the missing frames from the surrounding frames.
This induction is useful for learning the time-frequency structure in a mel-spectrogram in a self-supervised manner.
Note that similar effects have been observed for similar tasks in other fields (e.g., image inpainting~\cite{DPathakCVPR2016} and text infilling~\cite{WFedusICLR2018,JDevlinNAACL2019}), as mentioned in Section~\ref{sec:introduction}.
Finally, it should be noted that
(1) unlike CycleGAN-VC3, which uses TFAN, MaskCycleGAN-VC does not need a large increase in the converter parameters (only the input channels are doubled to receive $\bm{m}$ along with $\hat{\bm{x}}$), and
(2) FIF is a type of self-supervised learning; therefore, neither extra data nor a pretrained model (e.g., linguistic information) is required.

\subsection{Conversion with all-ones mask}
\label{subsec:conversion}

As a remaining question, what mask should be used during the conversion process (i.e., test phase)?
For this question, we simply use an all-ones mask.
Thus, we can convert speech under the assumption that no missing frames exist.
This assumption is the same as that used in typical VC.

\section{Experiments}
\label{sec:experiments}

\subsection{Experimental conditions}
\label{subsec:experimental_conditions}

\textbf{Dataset.}
We examined the effectiveness of MaskCycleGAN-VC on the Spoke (i.e., non-parallel VC) task of VCC 2018~\cite{VCC2018}, which contains recordings of native speakers of American English.
We used a subset of speakers that covers all inter- and intra-gender VC, i.e., VCC2SF3 (\textit{SF}), VCC2SM3 (\textit{SM}), VCC2TF1 (\textit{TF}), and VCC2TM1 (\textit{TM}), where \textit{S}, \textit{T}, \textit{F}, and \textit{M} indicate the sources, targets, females, and males, respectively.
We used combinations of 2 sources $\times$ 2 targets for the evaluation.
For each speaker, we used 81 sentences for training (of approximately 5 min in length, which is relatively short for VC) and 35 sentences for the evaluation.
Note that the training set contains no overlapping utterances between the source and target speakers; therefore, we need to train a converter in a fully non-parallel setting.
In this dataset, audio clips were down-sampled to 22.05 kHz.
Similar to a study on CycleGAN-VC3~\cite{TKanekoIS2020}, we extracted an 80-dimensional log mel-spectrogram with a window length of 1024 and hop length of 256 samples.

\smallskip\noindent\textbf{Conversion and synthesis process.}
For a fair comparison with CycleGAN-VC3~\cite{TKanekoIS2020}, we used the same conversion and synthesis process as CycleGAN-VC3.
Namely, we applied MaskCycleGAN-VC to mel-spectrogram conversion and synthesized the waveform using the pretrained MelGAN vocoder~\cite{KKumarNeurIPS2019}.\footnote{\url{https://github.com/descriptinc/melgan-neurips}}
Although for a fair comparison we did not change the parameters of the vocoder, fine-tuning it for each speaker is acceptable.

\smallskip\noindent\textbf{Network architectures.}
We used similar network architectures as in CycleGAN-VC2 for mel-spectrogram conversion, which was used as the baseline in the study on CycleGAN-VC3~\cite{TKanekoIS2020} (see Figure~4 in \cite{TKanekoICASSP2019} and Section 4.1 in \cite{TKanekoIS2020} for the details).
The converter consists of a 2-1-2D CNN~\cite{TKanekoICASSP2019}, and the discriminator is PatchGAN~\cite{CLiECCV2016}.
As mentioned in Section~\ref{subsec:fif}, the only difference between CycleGAN-VC2 and MaskCycleGAN-VC is that the input channels are doubled in the converter to receive $\bm{m}$ along with $\hat{\bm{x}}$.

\smallskip\noindent\textbf{Training settings.}
We used the same training settings as in CycleGAN-VC3~\cite{TKanekoIS2020}.
During the preprocessing, we normalized the mel-spectrograms using the training set statistics.
We used a least-squares GAN~\cite{XMaoICCV2017} as the GAN objective.
We trained the networks for $500k$ iterations using an Adam optimizer~\cite{DPKingmaICLR2015}, with the learning rates of the converter and discriminator set to 0.0002 and 0.0001, respectively, under momentum terms $\beta_1$ and $\beta_2$ of 0.5 and 0.999, respectively.
The batch size was set to 1, where each training sample consisted of 64 randomly cropped frames (approximately 0.75 s in length).
$\lambda_{cyc}$ and $\lambda_{id}$ were set to 10 and 5, respectively, and $\mathcal{L}_{id}$ was used for only the first $10k$ iterations to prevent $\mathcal{L}_{id}$ from disturbing the learning of conversion.
Similar to the previous CycleGAN-VCs, \textit{we did not use extra data, pretrained models, or a time alignment procedure for training}.

\subsection{Objective evaluation}
\label{subsec:objective_evaluation}

We conducted an objective evaluation to examine the differences in performance when using different components.
Because a direct comparison between the converted and target mel-spectrograms is difficult owing to the lack of a correct alignment, we used two metrics:
(1) \textit{mel-cepstral distortion (MCD)}, which is the most commonly applied measure and calculates the distance within the mel-cepstral domain (particularly, a 35-dimensional mel-cepstrum was extracted from the converted or targeted waveform using the WORLD analyzer~\cite{MMoriseIEICE2016}), and
(2) \textit{Kernel DeepSpeech Distance (KDSD)}~\cite{MBinkowskiICLR2020}, which computes the maximum mean discrepancy within the DeepSpeech2 feature space~\cite{DAmodeiICML2016} and is shown to be well correlated with human judgement~\cite{MBinkowskiICLR2020}.
For both metrics, the smaller the value, the better the performance.

\begin{table}[t]
  \vspace{-3.5mm}
  \caption{Comparison of MCD and KDSD using (a) different-sized masks, (b) different types of masks, and (c) different CycleGAN-VCs.
    The results are listed as MCD [dB]/KDSD [$\times 10^5$].
    Bold numbers indicate the best scores.
    \#param represents the number of converter parameters.}
  \label{tab:objective_evaluation}
  \vspace{1mm}
  \centering
  \scriptsize{
  \begin{tabularx}{\columnwidth}{clCCCCC}
    \toprule
    No. & (a) Size & SF-TF & SM-TM & SF-TM & SM-TF & \#param
    \\ \midrule
    1 & FIF 0
    & 7.66/786
    & 7.11/356
    & 6.91/277
    & 8.11/1094
    & 16M
    \\
    2 & FIF 25
    & 7.45/560
    & 6.85/297
    & 6.76/249
    & 7.84/775
    & 16M
    \\ \midrule
    3 & FIF 0-25
    & 7.45/489
    & 6.83/103
    & 6.78/206
    & 7.80/605
    & 16M
    \\
    4 & FIF 0-50
    & \textbf{7.37}/\textbf{467}
    & 6.77/\textbf{83.8}
    & 6.73/\textbf{146}
    & \textbf{7.64}/\textbf{502}
    & 16M
    \\
    5 & FIF 0-75
    & 7.40/468
    & \textbf{6.75}/89.2
    & \textbf{6.72}/169
    & 7.66/546
    & 16M
    \\ \bottomrule
    \toprule
    No. & (b) Type & SF-TF & SM-TM & SF-TM & SM-TF & \#param
    \\ \midrule
    6 & FIF
    & \textbf{7.37}/\textbf{467}
    & \textbf{6.77}/\textbf{83.8}
    & \textbf{6.73}/\textbf{146}
    & \textbf{7.64}/\textbf{502}
    & 16M
    \\
    7 & FIF$_{\text{NS}}$
    & 7.53/648
    & 7.00/638
    & 6.90/270
    & 7.97/1181
    & 16M
    \\
    8 & FIS
    & 7.52/727
    & 6.95/437
    & 6.88/418
    & 7.94/974
    & 16M
    \\
    9 & FIP
    & 7.65/920
    & 6.97/449
    & 7.09/774
    & 8.24/2126
    & 16M
    \\ \bottomrule
    \toprule
    No. & (c) Model & SF-TF & SM-TM & SF-TM & SM-TF & \#param
    \\ \midrule
    10 & Mask
    & \textbf{7.37}/467
    & \textbf{6.77}/\textbf{83.8}
    & \textbf{6.73}/\textbf{146}
    & \textbf{7.64}/\textbf{502}
    & 16M
    \\
    11 & V2
    & 7.66/891
    & 7.07/509
    & 6.96/494
    & 8.07/1107
    & 16M
    \\
    12 & V3
    & 7.54/\textbf{369}
    & 7.10/227
    & 6.91/311
    & 7.97/819
    & 27M
    \\ \bottomrule
    \end{tabularx}
  }
  \vspace{-3.5mm}
\end{table}

\smallskip\noindent\textbf{Comparison among different-sized masks.}
We first examined the effect of the mask size selection.
Here, the mask size indicates the size of the zero region (i.e., the black region in Fig.~\ref{fig:pipeline}).
We tested two variations.
(1) \textit{FIF X}: The mask size is constantly $X \%$ (i.e., $64 \times \frac{X}{100}$ frames). Here, \textit{FIF 0} means that an all-ones mask is used.
(2) \textit{FIF 0-X}: The mask size is randomly determined within the range of $[0, X \%]$. We list the results in Table~\ref{tab:objective_evaluation}(a).
We found that
(i) FIF with a non-zero-sized mask (Nos. 2--5) outperformed that with a zero-sized mask (No. 1) regardless of the mask size,
(ii) the performance is affected by the mask size (Nos. 3--5) and maximizes at approximately $X = 50$, and
(iii) FIF with a random-sized mask (No. 4) outperformed FIF with a constant-sized mask (No. 2) despite the same average size.
The possible reason is that, during training, the former includes an all-ones mask, which is used in the test phase, whereas the latter does not.

\smallskip\noindent\textbf{Comparison among different types of masks.}
We inspect the effect of the mask type selection.
We compared four variations.
(1) \textit{FIF}: Subsequent frames are masked, as shown in Fig.~\ref{fig:pipeline}.
(2) \textit{FIF$_{\text{NS}}$}: Non-subsequent frames (i.e., each frame is independently and randomly selected) are masked.
(3) \textit{FIS}: Subsequent spectrum bands (e.g., 45th--60th mel-spectrograms) are masked.
(4) \textit{FIP}: Mel-spectrogram was masked in a point-wise manner similar to a dropout~\cite{NSrivastavaJMLR2014}.
Under all settings, we used a mask size of \textit{0-50}, which was the best setting in the previous experiment.
We summarize the results in Table~\ref{tab:objective_evaluation}(b).
We found that \textit{FIF} (No. 6) outperformed the others (Nos. 7--9) for all speaker pairs.
We consider that, although learning the temporal structure is the most difficult, it is important for CycleGAN-VC2, and \textit{FIF} is the most effective in mitigating this difficulty.

\smallskip\noindent\textbf{Comparison among CycleGAN-VCs.}
We examined the differences in performance among
(1) MaskCycleGAN-VC (\textit{Mask}, particularly \textit{FIF 0-50}, was used);
(2) CycleGAN-VC2~\cite{TKanekoICASSP2019} (\textit{V2}), which was the same as \textit{Mask} except FIF was not used; and
(3) CycleGAN-VC3~\cite{TKanekoIS2020} (\textit{V3}), which applied TFAN instead of FIF.
The results are listed in Table~\ref{tab:objective_evaluation}(c).
We found that \textit{Mask} (No. 10) outperformed both \textit{V2} (No. 11) and \textit{V3} (No. 12) in most cases, reducing the model size compared to \textit{V3}.
Further evidence is provided in the next section.

\begin{figure}[t]
  \centerline{\includegraphics[width=\columnwidth]{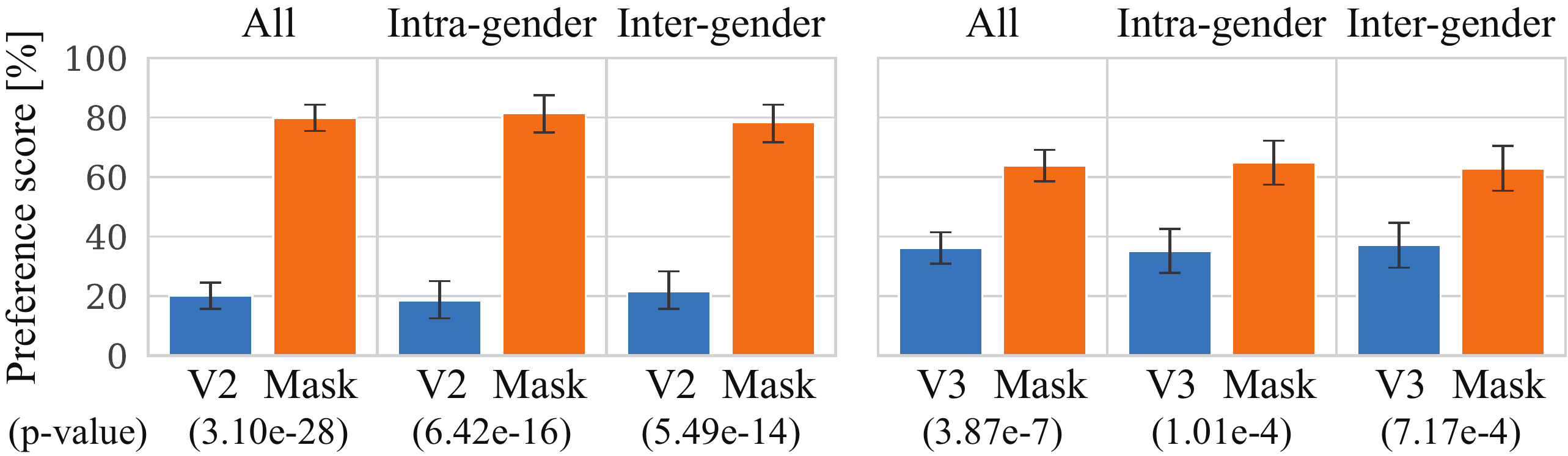}}
  \vspace{-3mm}
  \caption{Average preference scores on naturalness with $95\%$ confidence intervals.
    The numbers in parentheses indicate the p-values computed using a one-tailed binomial test.}
  \label{fig:ab_naturalness}
\end{figure}

\begin{figure}[t]
  \centerline{\includegraphics[width=\columnwidth]{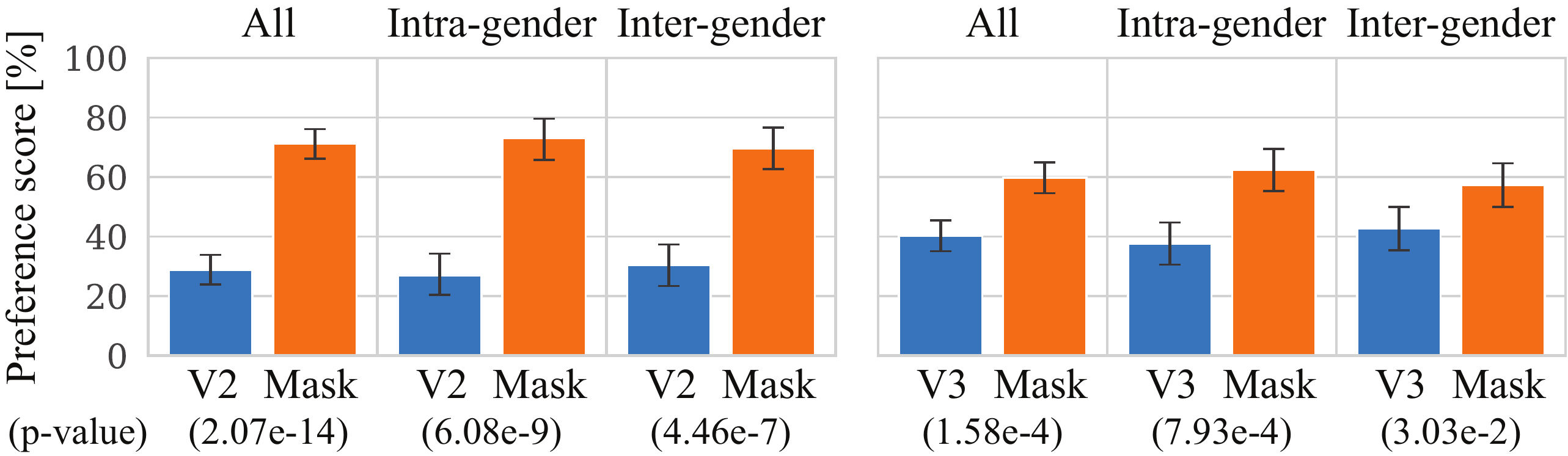}}
  \vspace{-3mm}
  \caption{Average preference scores on speaker similarity with $95\%$ confidence intervals.
    The numbers in parentheses denote the p-values calculated using a one-tailed binomial test.}
  \vspace{-3.7mm}
  \label{fig:xab_similarity}
\end{figure}

\subsection{Subjective evaluation}
\label{subsec:subjective_evaluation}

We conducted listening tests to investigate the differences in perceptual quality.
As the benchmark performance of mel-spectrogram conversion based on CycleGAN-VCs was previously examined in~\cite{TKanekoIS2020}, we investigated the comparative performance between \textit{Mask} and \textit{V2} and that between \textit{Mask} and \textit{V3} using two forced-choice preference tests.
In the AB test on naturalness, each listener was presented with two speech samples (A and B) and asked to choose their preferred one (A or B) considering both naturalness and intelligibility.
In the XAB test on speaker similarity, each listener was presented with three speech samples, including comparison targets (A and B) and a reference with a different utterance (X), and asked to choose their preferred one (A or B) with speaker characteristics closer to that of X.
These tests were conducted online, and 15 and 16 listeners participated in the AB and XAB tests, respectively.
Sentences, comparison targets, and the compared order (AB or BA) were randomly chosen from the collection of speech samples.
We gathered at least 300 answers for each model pair.
Audio samples are available from the link\footnoteref{foot:samples} presented in the first page.

We show the results of the AB test on naturalness and the XAB test on speaker similarity in Figs.~\ref{fig:ab_naturalness} and \ref{fig:xab_similarity}, respectively.
We found that in both tests, \textit{Mask} achieved statistically significantly better scores than \textit{V2} and \textit{V3} with a p-value of $<$ 0.05 (where a one-tailed binomial test was used).

\section{Conclusions}
\label{sec:conclusions}

Motivated by recent advances in mel-spectrogram vocoders, we proposed MaskCycleGAN-VC, which is an improvement of CycleGAN-VC2 for mel-spectrogram conversion.
To learn the time-frequency structure in a mel-spectrogram without an additional module such as TFAN, we introduced FIF, which allows the converter to learn such a structure in a self-supervised manner.
The experimental results showed that MaskCycleGAN-VC outperformed both CycleGAN-VC2 and CycleGAN-VC3 while maintaining a model size similar to that of CycleGAN-VC2.
Examining the generality of FIF is an interesting research topic, and future work includes applications to multi-domain VC~\cite{HKameokaSLT2018,TKanekoIS2019,HKameokaTASLP2020b} and application-side VC~\cite{AKainSC2007,KNakamuraSC2012,ZInanogluSC2009,TTodaTASLP2012,DFelpsSC2009,TKanekoIS2017b}.

\smallskip\noindent
{\bf Acknowledgements:}
This work was supported by JSPS KAKENHI 17H01763 and JST CREST Grant Number JPMJCR19A3, Japan.

\vfill\pagebreak

\bibliographystyle{IEEEbib}
\scriptsize{\bibliography{refs}}

\end{document}